# Observation of Cavity-Mediated Nonlinear Landau Fan and Modified Landau Level Degeneracy in Graphene Quantum Transport

Hongxia Xue[1,4]†, Hsun-Chi Chan[1,5]†, Zuzhang Lin[1,5]†, Dalin Boriçi[3]†, Shaobo Zhou[1,5]†, Yanan Wang[1,5], Kenji Watanabe[2], Takashi Taniguchi[2], Cristiano Ciuti[3]\*, Wang Yao[1,4,5], Dong-Keun Ki[1,4]\*, Shuang Zhang[1,5,6]\*

[1]Department of Physics, The University of Hong Kong, Pokfulam Road, Hong Kong, China.
[2]National Institute for Materials Science; Namiki 1-1, Tsukuba, 305-0044, Ibaraki, Japan.
[3]Université Paris Cité, CNRS, Matériaux et Phénomènes Quantiques, Paris, France
[4]HK Institute of Quantum Science & Technology, The University of Hong Kong, Pokfulam Road, Hong Kong, China.
[5]New Cornerstone Science Laboratory, University of Hong Kong, Hong Kong, China.
[6]Department of Electrical & Electronic Engineering; University of Hong Kong, Hong Kong, China.

\*Corresponding authors. Email: cristiano.ciuti@u-paris.fr; dkki@hku.hk; shuzhang@hku.hk

†These authors contributed equally to this work.

**Abstract:** Recent studies on cavity-coupled two-dimensional electron gas demonstrate that vacuum-field engineering can tailor electronic transport properties of materials. By achieving ultra-strong coupling between a terahertz resonator and mesoscopic graphene, we demonstrate that cavity vacuum fields can alter the effective degeneracies of Landau levels, resulting in a nonlinear Landau fan diagram for massless Dirac fermions while preserving quantum-Hall quantization. Specifically, by leveraging graphene's gate-tunability, we observe that quantum-Hall features—minimum longitudinal and quantized Hall conductance for a filling factor $\nu$, $\sigma_{xx} \approx 0$ and $\sigma_{xy} = \nu e^2/h$, respectively—occur at carrier densities reduced by more than 20% ($n < \nu eB/h$) compared to systems without cavity. This deviation increases with higher magnetic field (B) and $\nu$. Theoretical analysis attributes this effect to the virtual cavity photon mediated transitions between the non-equidistant Landau levels in graphene, significantly reducing their effective degeneracy. This study paves the way for investigating cavity quantum electrodynamics in highly tunable, atomically thin two-dimensional crystals.

Cavity quantum electrodynamics (cQED) aims to investigate and control the photon-dressed quantum states in the systems with well-defined energy levels, such as atoms and molecules, by coupling them with the discrete photon modes in the cavity (*1, 2*). Due to the non-trivial topology and flat band nature, Landau levels (LLs) in two-dimensional electron gas (2DEG) have become an ideal platform for exploring the cQED at a mesoscopic scale beyond atoms or molecules even in vacuum, *i.e.*, without any optical illumination (*1-7*). Experiments on 2DEG in GaAs/AlGaAs heterostructures have revealed that the strong cavity vacuum fields can mediate long-range electron hopping, enabling inter-edge coupling that disrupts the topological protection of the integer quantum-Hall effect (*5*). Recently, by hovering ring resonators over 2DEG, the vacuum fields were further tuned to reduce the exchange splitting and enlarge the fractional quantum-Hall gaps (*7*). Additionally, exotic phases like cavity-mediated superconductivity (*8*), superfluidity (*9*), and topological states (*2, 10-13*) have been predicted theoretically for various types of low-dimensional systems.

Graphene is a unique 2DEG that exhibits a linear dispersion relation at low energy, where the charge carriers behave as massless Dirac fermions (*14-18*). Over the past two decades, many studies have proven the relativistic nature of charge transport in graphene (*16-19*). In particular, under perpendicular magnetic fields ($B$), graphene exhibits quantum-Hall features from non-equidistant LLs of massless fermions with $E_N = \text{sgn}(N) v_F \sqrt{2e\hbar|N|B} \approx \text{sgn}(N) \times 36.28\sqrt{|N|B}$ meV, where $v_F \simeq 10^6$ m/s is the Fermi velocity, $N$ is a LL index in electron ($N \geq 0$) and hole ($N \leq 0$) sides (*14, 15*). Since the non-equidistant LLs impose distinct selection rules for transitions between LLs from those for equidistant LLs in conventional 2DEG (Figs. S1A and B), the cQED experiments in graphene could uniquely allow us to explore quantum transport and emergent collective excitations of the photon-dressed massless Dirac fermions.

In this work, by coupling a terahertz (THz) cavity with graphene (Figs. 1A, B), we demonstrate that the cavity vacuum fields can strongly modify quantum transport of massless Dirac fermions in graphene, particularly reducing the effective degeneracies of LLs, defined as $D_{eff} \equiv n/\nu$ for a given $\sigma_{xy} = \nu e^2/h$, from the well-established value of $eB/h$, leading to non-linear Landau fan diagram at finite magnetic fields as shown in Figs. 1 and 2. We found that the quantum-Hall features—both minima in longitudinal conductance and plateaus in Hall conductance for a filling factor $\nu$, $\sigma_{xx} \approx 0$ and $\sigma_{xy} = \nu e^2/h$, respectively—occur at lower carrier density ($n < \nu eB/h \rightarrow D_{eff} = n/\nu < eB/h$) as increasing $B$ and $\nu$, with a reduction factor exceeding 20%. The effect could be reproduced in different samples (Supplementary Materials) and is different from the previous results from the cavity-coupled 2DEG in GaAs/AlGaAs (*5-7*). Our theoretical models reveal that the non-equidistant LLs of graphene, combined with enriched selection rules allowing a large number of interband transitions - unlike those in the conventional 2DEG - drive this effect (Fig. 3). This study not only proves the unique nature of the photon-dressed massless Dirac fermions but also opens new possibilities to explore cQED in a variety of 2D materials and heterostructures with large experimental flexibilities (*20-25*).

**Ultra-strong coupling between C$_4$ cavity and graphene**

To realize ultra-strong coupling between cavity and graphene, we designed a metallic resonator with a C$_4$ rotational symmetry and a resonant frequency of $f_0 \sim 4$ THz ($hf_0 \sim 16.54$ meV), as shown in Fig. 1A. This antenna-like cavity features a central hot spot with the tunable arm length for precise resonant frequency control, enabling strong coupling in recent studies (*26, 27*). The gap between the metallic rods is around 2.5 μm within which the vacuum field fluctuations near $f_0$ is significantly enhanced (Fig. 1B). The hBN/graphene/hBN square sample with four electrical contacts in a van der Pauw (vdP) geometry were prepared by van der Waals (vdW) assembly and conventional nanofabrication methods (see Supplementary Materials) at the center of the resonator gap to be fully immersed in the focused vacuum fields (see sample images in the bottom inset of Fig. 1C). The simulated transmission color map (see Supplementary Materials) shows an anti-crossing behavior with a strong vacuum Rabi splitting at $\Omega_R/\omega_0 \sim 0.1$ ($\omega_0 = 2\pi f_0$), indicating that the system operates in the ultra-strong coupling regime (*28, 29*). The reference sample without cavity was also prepared for a direct comparison (see the bottom inset of Fig. 1C). The electrical measurements were carried out at low-temperature cryostats with superconducting magnets using a vdP method (see Supplementary Materials and Fig. S2 for details).

**Observation of nonlinear Landau fan diagram**

The main experimental findings of this work—the nonlinear Landau fan diagram and significantly reduced Landau level degeneracies —are summarized in Figs. 1C and 2. In Fig.

1C, we plot the longitudinal and Hall conductivities, $\sigma_{xx}$ and $\sigma_{xy}$ respectively, as a function of $n$ measured at $B = 9$ T in the quantum-Hall regime (see Supplementary Materials for details of calculating $n$). Interestingly, while both the reference and C$_4$ cavity samples show clear minima in $\sigma_{xx}$ and fully developed plateaus in $\sigma_{xy}$ quantized at $\nu e^2/h$ with $\nu = \pm 2, \pm 6, \pm 10, \ldots$, the characteristic half-integer quantum-Hall effect in graphene (*14, 15*), the quantum-Hall features from the cavity sample shifted to lower $n$ compared to those from the reference (see, *e.g.*, the orange bands for the features for $\nu = -10$). Furthermore, near zero density, the features from the cavity and the reference samples merge well with each other as shown in the top inset of Fig. 1C. This indicates that in the cavity-coupled graphene, the quantum-Hall features no longer occur at a constant LL filling factor $\nu_n \equiv \frac{nh}{eB} \neq \pm 2, \pm 6, \pm 10, \ldots$, but appear at lower $n < \nu eB/h$ at fixed $B$ (to avoid confusion, we will use $\nu$ to identify the value of $\sigma_{xy} = \nu e^2/h$ and $\nu_n$ to discuss the corresponding position in $n$ for a given $B$).

For more comprehensive analysis, we measured the Landau fan diagram, $R_{xx}(V_g, B)$ and $R_{xy}(V_g, B)$, of both the cavity and reference sample by sweeping a back gate voltage ($V_g$) to tune $n$ at different $B$ values. As shown in Figs. 2A and S3, the cavity sample exhibits the minima in $R_{xx}$ (the dark bands in Fig. 2A) and plateaus in $R_{xy}$ (the bands with the same colors in Fig. S3B) that fan out simultaneously from the charge neutrality point at zero $B$. However, we find that they do not follow the linear lines corresponding to the points in $(V_g, B)$ with a fixed $\nu_n$ (the broken lines in Fig. 2A) as expected for the quantum-Hall features that should appear at a fixed LL filling, $\nu_n \propto V_g/B$, but they are bent inwards toward zero density as increasing $B$. On the other hand, at low $B$, the features follow the linear lines well as shown in Fig. 2B. This is consistent with the shift of the quantum-Hall features found in Fig. 1C at 9 T. Notably, we could not find a single capacitance value that can force the quantum-Hall features to follow the set of linear lines in the Landau fan diagram across the entire $(V_g, B)$ map. In contrast, the reference sample (Fig. 2C) exhibits a linear Landau fan diagram in all $V_g$ and $B$ ranges investigated following the integer $\nu_n = \nu = \pm 2, \pm 6, \pm 10, \ldots$, as expected. This signifies the effect of the cavity on the non-linear Landau fan diagram in the cavity-coupled graphene, suggesting significantly reduced degeneracies of LLs at large $B$ and $n$.

For more direct comparison, we plot the positions of the minima $R_{xx}$ shown in Fig. 2A against the linear lines corresponding to the integer $\nu_n$ in Fig. 2D that exhibits clear shifts of the minima $R_{xx}$ with respect to the linear lines. From this, we can further extract the amount of the shift in density, $\Delta n \equiv n_m - \nu eB/h$ ($n_m$: the density at the minimum $R_{xx}$), and plot them as a function of $B$ for different integer $\nu$ in Fig. 2E. Both plots clearly show that the quantum-Hall features in the cavity sample exhibit a nonlinear behavior in $(V_g, B)$ deviating from the linear lines more at larger $B$ and $\nu$. In addition, Fig. 2E shows the $\Delta n$ is as large as ~$10^{12}$ cm$^{-2}$, corresponding to ~13 V change in $V_g$ that is beyond any possible errors in determining the charge neutrality point and/or the gate capacitance used to calculate $n$. Moreover, as discussed in Supplementary Materials, we can rule out any known possible scenarios that may cause the bending, such as the electric-field focusing near the sample edge (*30*) or the presence of the open orbits in graphene moiré systems (*31*). The features are also reproduced in other samples as shown in Figs. S5-7.

**Discussions and theoretical analysis**

The Landau fan diagram displays quantum-Hall features in gate voltage (*i.e.*, carrier density) and magnetic field, depending solely on how the LLs are filled not their dispersion in energy. Thus, regardless of the LL dispersion, the Landau fan diagram should always exhibit linear lines. Previous studies on the cavity-coupled 2DEG in GaAs/AlGaAs reported either quantum-Hall features appearing at lower $B$ than the reference sample at a similar $n$, corresponding to a higher $\nu$ (opposite to ours) (*5*), or no obvious change in the positions of the quantum-Hall features in $B$ between the sample with and without cavity (*6, 7*). In 2DEG systems where the Landau levels are equally spaced and bosonization is rigorous, a straightforward calculation based on Bogliubov-Hopfield transformation can rigorously show that the quantum Hall features remain unchanged regardless of the coupling strength between the cavity field and the Landau transitions. Hence any modification in the case of cavity-couped 2DEG likely arise from cavity mediated coupling between disorder modes or edge modes of a finite sample (*3, 32*).

Our findings indicate that, in addition to the cavity, the non-equidistant LLs and the allowed interband transitions in graphene also play significant roles in influencing the LL filling under cavity vacuum fields. As illustrated in Fig. 3A, for a realistic magnetic field strength, many LLs exist on both the electron and hole sides within the linear dispersion regime of graphene's Dirac cone. Besides intraband transitions, the selection rule allows a large number of interband transitions. For instance, at $B = 9T$, the linear dispersion regime (approximately $\pm 1eV$ around the Dirac point) can accommodate approximately 90 LLs on each side of the Dirac point, leading to approximately 180 interband transitions. These interband transitions all couple to the same cavity mode, leading to collectively coherent interactions that strongly modify the quantum transport. This is in sharp contrast to the case of cavity coupled 2DEG where only a single intraband transition couples to the cavity.

In addition, we were able to find similar nonlinear behavior in other graphene samples with different cavity geometries, such as double- and single-ring resonators (Figs. S6 and S7). Even though the details, such as the electron-hole asymmetry and the size of the shift, vary from sample to sample, this suggests that the specific cavity geometry does not affect the emergence of the non-linear Landau fan diagram significantly. Thus, to capture the key aspects of the phenomenon, we can begin with a more general model without considering the specific cavity geometry in detail.

To explain the experimental observation, we perform exact diagonalization (ED) calculations on the cavity-coupled graphene system and use the Kubo formula (*33*) to calculate the Hall conductivity when electrons fill up to the $N$-th LL. We work in the Landau gauge and consider only transitions that conserve the wave vector k. This approximation is valid in the absence of electronic disorder, edge potentials, or spatial gradients in the cavity mode field. The effect we are investigating is thus an intrinsic bulk property. This simplification allows for exact diagonalization, as it reduces the problem to a number of fermionic states equal to the number of Landau levels. For the cavity coupled graphene system, the total Hamiltonian reads

$$\widehat{H}_{tot} = \hbar\omega \hat{a}^\dagger \hat{a} + \widehat{H}_e + g(\hat{a}^\dagger + \hat{a})\widehat{H}_i. \quad (1)$$

Here, from left to right, the first term represents Hamiltonian for the cavity photons with frequency $\omega = 2\pi f_0$ and photon annihilation (creation) operator $\hat{a}$ ($\hat{a}^\dagger$), the second term $\widehat{H}_e = v_F\sqrt{2e\hbar B}\sum_{N=-\infty}^{\infty} \lambda\sqrt{|N|}\,\hat{c}_N^\dagger \hat{c}_N$ describes the graphene LLs with $\lambda \equiv \text{sgn}(N)$, the fermion annihilation (creation) operator $\hat{c}_N(\hat{c}_N^\dagger)$, and the last term denotes the cavity-graphene coupling from right- and left-handed polarization (RCP, LCP) transitions from $N$ to $N'$ LLs, where $g$ is the coupling constant and $\widehat{H}_i = \sum_{N,N'=-\infty}^{\infty}\left(\lambda C_{N'}^- C_N^+ \delta_{|N'|-1,|N|} + \lambda' C_{N'}^+ C_N^- \delta_{|N'|,|N|-1}\right)\hat{c}_{N'}^\dagger \hat{c}_N \equiv$

$\hat{J}_r + \hat{J}_l$ with $\hat{J}_r$ ($\hat{J}_l$) the current operator for RCP (LCP) transitions that obey the selection rule $|N'| - |N| = 1$ ($|N'| - |N| = -1$) (see section 6 in Supplementary Materials for derivations). We exactly solve the eigenvalue equation for $\hat{H}_{tot}$ and get a set of hybridized Landau polariton energy levels, $|E_i'\rangle$ with $i = 0, 1, 2, ...$, and their corresponding eigenfunctions. Applying the Kubo formula (30) to a single-degeneracy case, we find the calculated $\sigma_{xy}$ exhibits a significant enhancement (~18%) compared to the cavity-free system. To ensure the quantization of the overall $\sigma_{xy}$, the degeneracy of LLs must be modified to a lower value. In other words, due to the cavity, the $D_{eff} = n/\nu$ of graphene LLs becomes smaller than $eB/h$, making the overall slope of $\sigma_{xy}(n)$, $\delta\sigma_{xy}/\delta n \propto 1/D_{eff}$, larger than $e/B$ when there is a cavity. This aligns well with our observations in all cavity samples as shown in Figs. 1C and S5-S7. In contrast, Kubo bulk calculations with cavity in GaAs, which has only the intraband transitions (see Fig. 3B), show no change of Landau degeneracy (3,32). By using a perturbation approach, we can include more LLs within an energy cut-off of 1eV in the linear regime of graphene to calculate the carrier differences $\Delta n$ (see Figs. 2E and 3C) at $B = 9T$, which are in reasonable agreement with the experimental results. The discrepancy between ED calculations and experimental data may arise from the negligence of contributions from transitions in graphene's nonlinear dispersion region (i.e. beyond the $\pm 1eV$ linear range around the Dirac point), and that only a single optical mode of the cavity was considered (whereas the cavity hosts two degenerate optical modes). In addition to the ED calculations based on Kubo formula, we obtain consistent results using an independent theoretical approach based on quantum transport formalism (*34, 35*), which employs an effective electronic Hamiltonian incorporating the renormalization of single-body Hamiltonian due to the emission and absorption of virtual cavity photons (*36*). Within this framework, we have computed the Landau level degeneracy modified by the cavity, as well as the longitudinal and Hall conductance in a finite-size Hall bar geometry.

Remarkably, as shown below, the degeneracy of the Landau levels is reduced due to dressing by cavity virtual photons. Importantly, the change in degeneracy is Landau-level dependent, leading to a nonlinear modification of the Landau fan diagram. Furthermore, in agreement with experimental observations, the slope of the Hall conductance versus carrier density increases in the presence of the cavity—a direct consequence of the cavity-induced alteration of Landau level degeneracy. It is also worth noting that finite-size effects introduce an electron–hole asymmetry, even in the absence of the cavity, which is further enhanced by the cavity coupling.

**Conclusion and outlook**

In conclusion, we experimentally demonstrate a nonlinear Landau fan diagram in graphene, mediated by cavity vacuum fluctuations in the ultra-strong coupling regime. Our theoretical modeling suggests that this arises from the photon-induced transitions between the non-equidistant LLs in graphene, leading to a significant reduction in Landau degeneracies at large *B* and *n*. Our study proves the unique transport properties of the photon-dressed massless Dirac fermions that can allow us to investigate the coupling between the photon (the massless bosons) and the massless fermions on a mesoscopic scale. In addition, recent advances in vdW transfer techniques enable precise manipulation and construction of diverse types of artificial heterostructures including moiré materials (*20-25, 37*). This further facilitates the design of novel cavity-coupled architectures with unprecedented flexibility and material compatibility. Theoretically, the non-equidistant LLs in graphene provides a unique platform to study cQED beyond the two-level systems (*38, 39*) as the transitions between these LLs cannot be exactly transformed to a single transition between the two levels unlike the equidistant LLs in conventional 2DEG (*3, 29*).

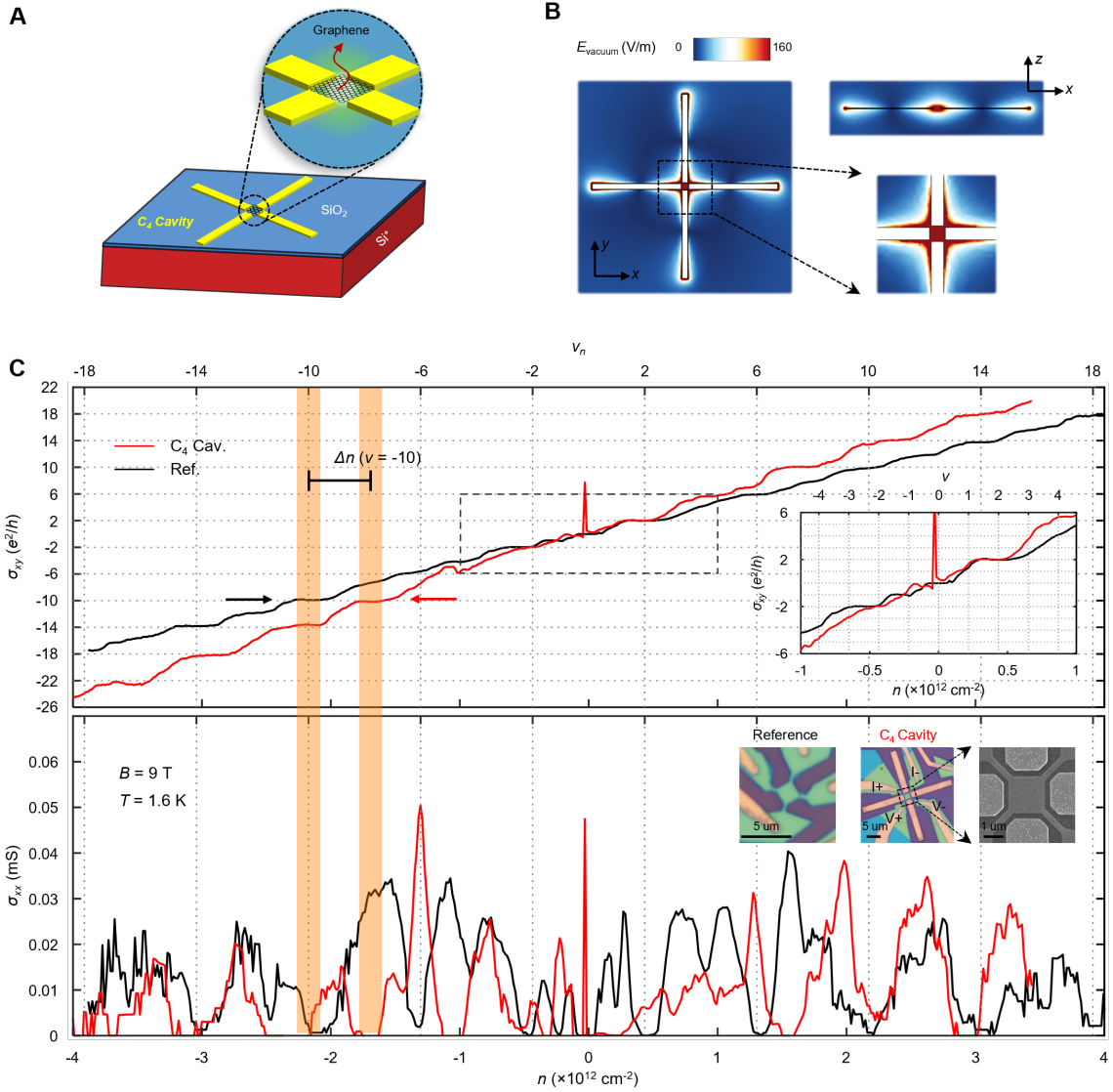

**Fig. 1. Graphene coupled with C₄ cavity resonator.** (**A**) Schematic diagram of graphene embedded in a cavity resonator with C₄ symmetry. (**B**) Top and side view of the simulated vacuum field distribution for the C₄ cavity. (**C**) Carrier density dependence of Hall (top, $\sigma_{xy}$) and longitudinal (bottom, $\sigma_{xx}$) conductance for graphene with and without C₄ cavity (red and black curves, respectively) measured at $B = 9$ T and $T = 1.6$ K. The top $x$-axis shows the LL filling factor $\nu_n$, calculated from $n$ and $B$ using the relation, $\nu_n \equiv \frac{nh}{eB}$. Both cavity-coupled and reference samples exhibit clear quantized $\sigma_{xy} = \nu\, e^2/h$ and minima in $\sigma_{xx}$ but the cavity-coupled sample (the red curves) exhibits the shift in $n$ as indicated by the black and red arrows at the top panel and the orange vertical bands for the $\nu = -10$ state. Top inset shows the magnified view of $\sigma_{xy}(n)$ near zero density, exhibiting an overlap between the black and red curves. The insets at the bottom show the optical and scanning electron microscopy images of the devices.

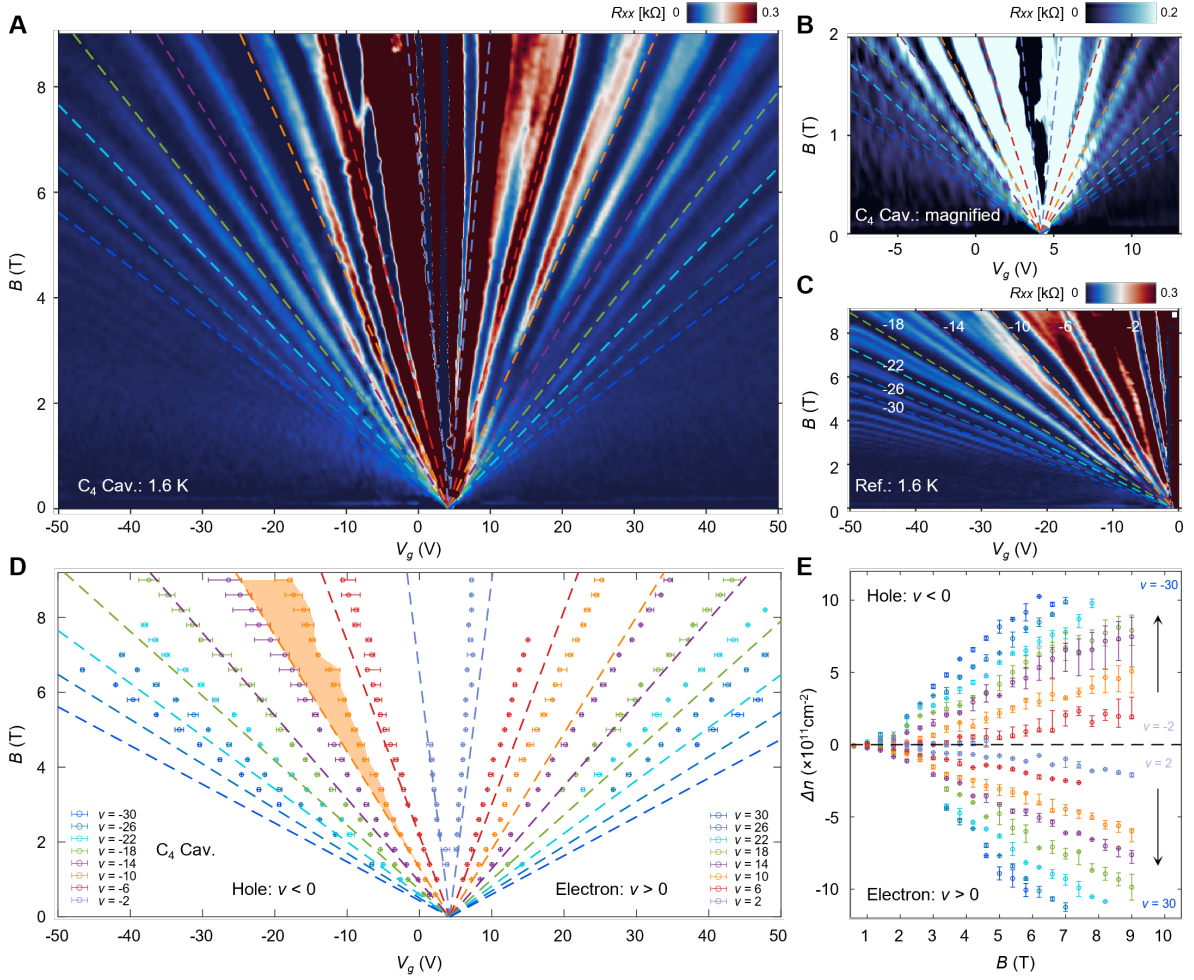

**Fig. 2. Nonlinear Landau fan diagram.** (**A**) Landau fan diagram—the color map of longitudinal resistance ($R_{xx}$) as a function of back-gate voltage ($V_g$) and magnetic field ($B$)— of the cavity-coupled graphene measured at $T = 1.6$ K. The linear dashed lines with different colors indicate the positions at which the LL filling factors are fixed at a set of integer values, $\nu_n = \pm 2, \pm 6, \pm 10, ...$, expected for graphene. The minima in $R_{xx}$ (dark blue regions) deviate from the linear lines more at higher $B$ and $\nu$, indicating its non-linear behavior. (**B**) The magnified view of the Landau fan diagram shown in (A) that exhibits linear behavior at small $B$. (**C**) Landau fan diagram of the reference sample without the cavity at $T = 1.6$ K, exhibiting linear behavior up to 9 T in all $V_g$ range investigated. The numbers indicate the filling factors of each dashed line. (**D**) Comparison between the circles that mark the positions of the $R_{xx}$ minima (the dark blue area) and the linear dashed lines in (A). Colors represent $\nu$ values. (**E**) Dependence of $\Delta n \equiv n_m - \nu eB/h$ ($n_m$: the density at the minimum $R_{xx}$) on $B$ for different $\nu$ values. The data points for $\nu = -10$ correspond to the orange shaded area in (D).

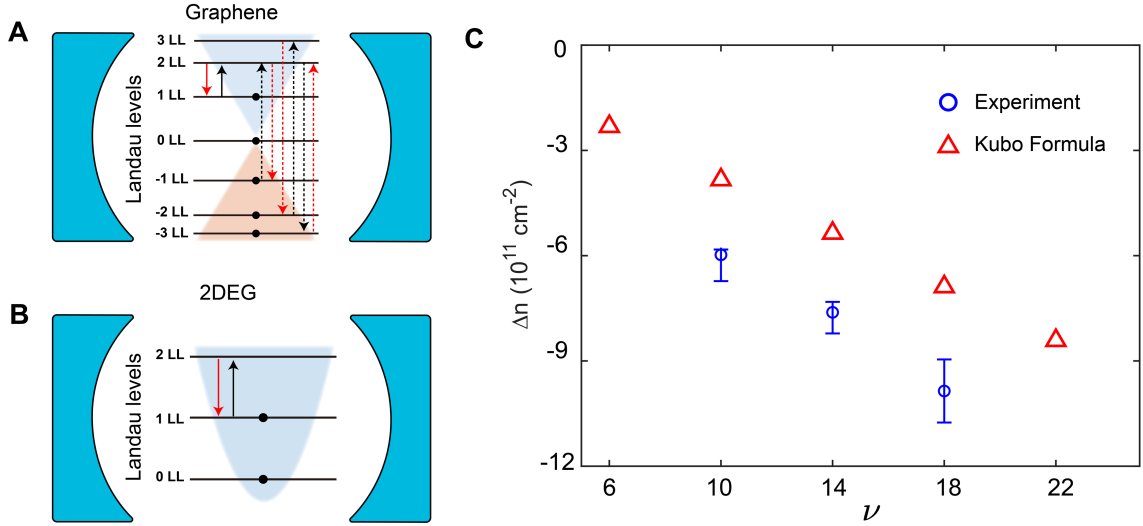

**Fig. 3. Vacuum-dressed Landau level degeneracies in graphene and 2DEG.** (**A**) Interband and intraband transitions among non-equidistant LLs in graphene in a cavity, which are dictated by the selection rules, yielding effective LL degeneracy $D_{eff} < eB/h$. (**B**) Intraband transitions between equidistant LLs in cavity-coupled 2DEG producing $D_{eff} = eB/h$ irrespective of the cavity coupling (in 2DEG, there is no interband transition). In Figs. 3A and 3B, the black dots and black (red) arrows indicate the occupied electron states and RCP (LCP) transitions between the LLs, respectively. (**C**) Dependence of $\Delta n$ on the electron side at $B = 9T$ for the ED calculations (red triangles) and the experimental data (blue circles with error bars, replotted from Fig. 2E).

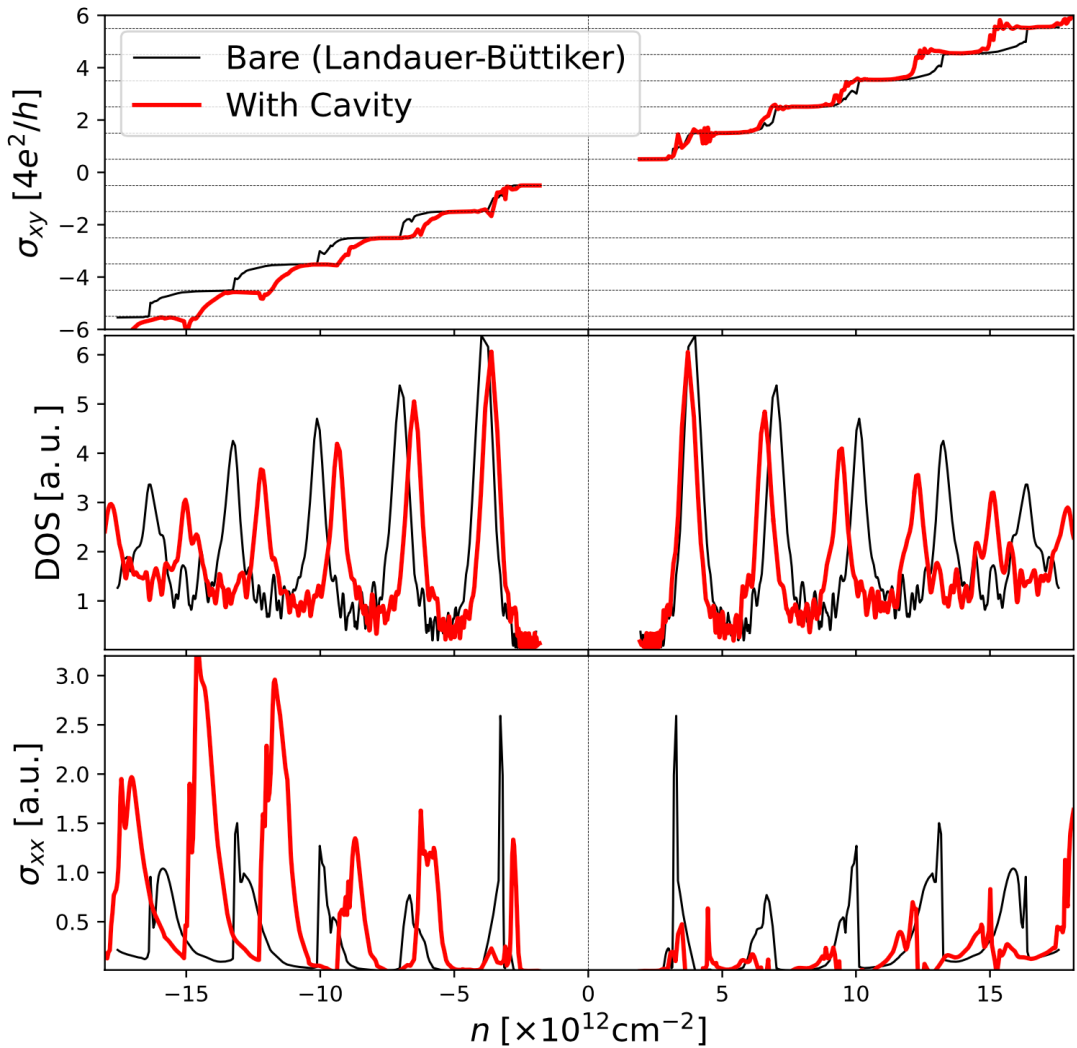

**Fig. 4. Cavity-modified quantum electron transport and density of states in graphene, computed using an effective electronic Hamiltonian where the cavity photon is adiabatically eliminated, combined with the Landauer-Büttiker formalism.** Top panel: Hall conductance as a function of carrier density. Middle panel: Density of states (DOS) at Fermi level versus carrier density. Bottom panel: Longitudinal conductance versus carrier density. Black curves correspond to the bare graphene system, while red curves include the effect of the cavity. The simulation is performed with a magnetic field of 20 Tesla and a six-terminal Hall bar of size 85 × 61 nanometers. Note that the magnetic field is stronger than in the experimental setup, and the system size is smaller than the actual sample to ensure numerical convergence and computational feasibility. Details are included in the Supplementary Material.

**Acknowledgments:** We thank Jingwen Ma, Huiyuan Zheng, Qiuchen Yan, and Tianyu Zhang for theoretical and technical support. We thank Hasebe Kazuki and Zhongfu Li for the discussion.

**Funding:** S.Z., D.K.K., W.Y., and C.C. acknowledge the financial support from a grant under the ANR/RGC Joint Research Scheme sponsored by Research Grants Council of Hong Kong and French National Research Agency (A-HKU705/21). S.Z., D.K.K., and W.Y. acknowledge supports from the Research Grants Council of Hong Kong under the Area of Excellence scheme, AoE/P701/20 (S.Z., D.K.K., W.Y.) and AoE/P-502/20 (S.Z.), and under the General Research Fund, GRF17309021 (S.Z.) and GRF17301424 (D.K.K.). S.Z. and W.Y. also acknowledge support from New Cornerstone Science Foundation. C.C. acknowledges support from French project TRIANGLE (ANR-20-CE47-001) and The National Research Agency (ANR) under the France 2030 program (ANR-24-RRII-0001). K.W. and T. T. acknowledge support from the JSPS KAKENHI (Grants No. 21H05233 and No. 23H02052) and World Premier International Research Center Initiative (WPI), MEXT, Japan.